\documentclass[12pt,thmsa]{article}
\usepackage{amsfonts}

\input{tcilatex}
\begin{document}

\author{Harald Grosse$^a$, Karl-Georg Schlesinger$^b$ \\
$^a$Institute for Theoretical Physics\\
University of Vienna\\
Boltzmanngasse 5\\
A-1090 Vienna, Austria\\
e-mail: grosse@doppler.thp.univie.ac.at\\
$^b$Institute for Theoretical Physics\\
University of Vienna\\
Boltzmanngasse 5\\
A-1090 Vienna, Austria\\
e-mail: kgschles@esi.ac.at}
\title{A remark on the motivic Galois group and the quantum coadjoint action}
\date{}
\maketitle

\begin{abstract}
It was suggested in \cite{Kon 1999} that the Grothendieck-Teichm\"{u}ller
group $GT$ should act on the Duflo isomorphism of $su(2)$ but the
corresponding realization of $GT$ turned out to be trivial. We show that a
solvable quotient of the motivic Galois group - which is supposed to agree
with $GT$ - is closely related to the quantum coadjoint action on $U_q\left(
sl_2\right) $ for $q$ a root of unity, i.e. in the quantum group case one
has a nontrivial realization of a quotient of the motivic Galois group. From
a discussion of the algebraic properties of this realization we conclude
that in more general cases than $U_q\left( sl_2\right) $ it should be
related to a quantum version of the motivic Galois group. Finally, we
discuss the relation of our construction to quantum field and string theory
and explain what we believe to be the ``physical reason'' behind this
relation between the motiviv Galois group and the quantum coadjoint action.
This might be a starting point for the generalization of our construction to
more invloved examples.
\end{abstract}

\section{Introduction}

In the seminal paper \cite{Kon 1999}, a symmetry on the space $\mathcal{D}%
\left( M\right) $ of deformation quantizations of a finite dimensional
manifold $M$, in the form of an action of a quotient of the motivic Galois
group (this quotient is supposed to be equivalent to the
Grothendieck-Teichm\"{u}ller group $GT$ as introduced in \cite{Dri}), is
discussed. Conjecturally, this is related to an action of $GT$ on the
extended moduli space (see \cite{Kon 1994}, \cite{Wit} for this notion) of
conformal field theories, as it appears in string theory. As a simple
example for the appearance of this symmetry, an action of $GT$ on the Duflo
isomorphism of finite dimensional Lie algebras is suggested. But it was
observed by Duflo (see \cite{Kon 1999}) that the corresponding realization
of $GT$ is trivial.

In this paper, we restrict the consideration to the case of the Lie algebra $%
su(2)$. We pose the question if a nontrivial realization of the motivic
Galois group can be observed in the $q$-deformed case. Since our argument
does not involve any $*$-structure, we can work with the quantum algebra $%
U_q\left( sl_2\right) $. We show that in the case where $q$ is a root of
unity, the quantum coadjoint action on $U_q\left( sl_2\right) $ (see \cite
{DK}, \cite{DKP}) is, indeed, closely related to a quotient of the motivic
Galois group which is stated in \cite{Kon 1999} to act on the Hochschild
cohomology of algebraic varieties and is conjectured, there, to be
equivalent to $GT$. It is precisely the much larger and highly nontrivial
center of $U_q\left( sl_2\right) $, appearing in the root of unity case,
which makes a nontrivial realization of the motivic Galois group possible.
This explains why one observes in the non-deformed case - but also in the
case of generic values of $q$ - only a trivial realization.

After collecting, in section 2, some basic background material on the
motivic Galois group and the quantum coadjoint action, we present, in
section 3, our construction. In section 4, we discuss the relation of our
result to quantum field and string theory. The algebraic structure appearing
in our construction suggests that in the case of more involved examples than 
$U_q\left( sl_2\right) $ a quantum version of the motivic Galois group will
appear.. We suggest a physical ``explanation'', rooted in properties of
quantum field theory, why our construction works, which might serve as a
starting point for the generalization to other examples. Section 5 contains
some concluding remarks.

We should utter a warning addressed to any potential reader of this paper
who is on a technical level acquainted with modern algebraic number theory:
The necessary specification of the precise type of motivic Galois group at
use (pro-nilpotent, pro-unipotent, etc.) is completely ignored in this
paper. We work with the (pro-unipotent) approach of \cite{Kon 1999} and feel
free to adopt the terminology ``the motivic Galois group'' used there.

\bigskip

\section{Background material on the motivic Galois group and the quantum
coadjoint action}

The Grothendieck-Teichm\"{u}ller group $GT$ is introduced in \cite{Dri} as a
kind of gauge freedom on the Drinfeld associator $\alpha $ and the $R$%
-matrix of any quasitriangular quasi-Hopf algebra (see \cite{Dri} for the
technical details). It was already observed there that the Lie algebra of $%
GT $ is closely related to the so called Ihara algebra (see \cite{Iha 1987}, 
\cite{Iha 1989}). The Ihara algebra has the following structure: Consider
formal expressions $\varphi \left( .\right) $ where the $\left( .\right) $
indicates that these expressions can be evaluated at any finite dimensional
metrizable - i.e. equipped with an invariant inner product - Lie algebra $g$%
. After evaluation, $\varphi \left( g\right) $ becomes an element of the
Poisson algebra defined by the Kirillov bracket $\left\{ ,\right\} _g$ of $g$%
. So, the $\varphi \left( .\right) $ are, roughly speaking, all universal
expressions which one can define for any finite dimensional metrizable Lie
algebra using the Kirillov bracket. The Ihara algebra is then defined as the
Lie algebra with bracket $\left[ ,\right] $ on the $\varphi \left( .\right) $
with 
\begin{equation}
\left[ \varphi _1,\varphi _2\right] \left( g\right) =\left\{ \varphi
_1\left( g\right) ,\varphi _2\left( g\right) \right\} _g  \label{1}
\end{equation}
It is conjectured in \cite{Kon 1999} that $GT$ can be identified with a
solvable quotient of the motivic Galois group the Lie algebra of this
quotient having generators $L_0,P_3,P_5,P_7,...$ and the bracket given by 
\begin{equation}
\left[ P_{2k+1},P_{2l+1}\right] =0  \label{2}
\end{equation}
and 
\begin{equation}
\left[ L_0,P_{2k+1}\right] =\left( 2k+1\right) P_{2k+1}  \label{3}
\end{equation}
for all $k,l\geq 1$. A proof is announced there that this quotient acts on
the Hochschild cohomology of any complex algebraic variety.

A simple explicit example for an action of $GT$ is suggested in \cite{Kon
1999} by considering the Duflo isomorphism of a finite dimensional Lie
algebra $g$ over $\Bbb{R}$: The Poincare-Birkhoff-Witt isomorphism gives a
linear isomorphism between the universal envelope $U\left( g\right) $ of $g$
and the algebra $Sym\left( g\right) $ of polynomials on $g^{*}$. This is,
obviously, not an algebra isomorphism since $U\left( g\right) $ is
noncommutative. But as shown in \cite{Duf}, the restriction of the
Poincare-Birkhoff-Witt map to a linear isomorphism between the center of $%
U\left( g\right) $ and the algebra $Sym\left( g\right) ^g$ of invariant
polynomials on $g^{*}$ becomes an algebra isomorphism after combining it
with an automorphism of $Sym\left( g\right) ^g$. This automorphism is
generated (see e.g. \cite{Kon 1997} for the technical details) by the formal
power series 
\begin{equation}
F\left( x\right) =\sqrt{\frac{e^{\frac x2}-e^{-\frac x2}}x}  \label{4}
\end{equation}
The claim is that there are other possible choices for $F$ than the
classical choice (\ref{4}) and that this ``gauge freedom'' of the Duflo
isomorphism is described by an action of $GT$. But this action of $GT$ turns
out to be trivial: It was observed by M. Duflo (see \cite{Kon 1999}) that,
though the different choices of $F$ generate different morphisms of $%
Sym\left( g\right) $, upon restriction to $Sym\left( g\right) ^g$ all
choices are equivalent.

We will show in this paper for the case $g=su(2)$ that upon passing to the
quantum algebra $U_q\left( sl_2\right) $ for $q$ a root of unity, one does
get a nontrivial action of the quotient of the motivic Galois group
determined by (\ref{2}) and (\ref{3}). Since we do not need any $*$%
-structure in our construction, we can work with the algebra $U_q\left(
sl_2\right) $ instead of a quantum version of $su(2)$. We will close this
section with a brief introduction to $U_q\left( sl_2\right) $ and the
quantum coadjoint action, and a short explanation why the question of an
action of the motivic Galois group related to the center of $U\left(
g\right) $, respectively, $U_q\left( g\right) $ is of special interest in
string theory.

\bigskip

The quantum algebra $U_q\left( sl_2\right) $, $q\in \Bbb{C}$ is defined as
the complex, associative, unital algebra with generators $e,f,k,k^{-1}$ and
relations 
\begin{eqnarray*}
kk^{-1} &=&k^{-1}k=1 \\
ke &=&q^2ek \\
kf &=&q^{-2}fk \\
\left[ e,f\right] &=&\frac{k-k^{-1}}{q-q^{-1}}
\end{eqnarray*}
In addition, $U_q\left( sl_2\right) $ carries a uniquely given Hopf algebra
structure with coproduct 
\begin{eqnarray*}
\Delta \left( e\right) &=&e\otimes 1+k\otimes e \\
\Delta \left( f\right) &=&f\otimes k^{-1}+1\otimes f \\
\Delta \left( k\right) &=&k\otimes k
\end{eqnarray*}
counit 
\begin{eqnarray*}
\varepsilon \left( e\right) &=&\varepsilon \left( f\right) =0 \\
\varepsilon \left( k\right) &=&1
\end{eqnarray*}
and antipode 
\begin{eqnarray*}
\Gamma \left( e\right) &=&-k^{-1}e \\
\Gamma \left( f\right) &=&-fk \\
\Gamma \left( k\right) &=&k^{-1}
\end{eqnarray*}
For $q$ not a root of unity, i.e. $q^n\neq 1$ for all $n\in \Bbb{N}$, the
center of $U_q\left( sl_2\right) $ is generated by the quantum Casimir $C_q$
given by 
\begin{equation}
C_q=qk+q^{-1}k^{-1}+\left( q-q^{-1}\right) ^2fe  \label{5}
\end{equation}
Now, let $q$ be a primitive $l$-th root of unity with $l\geq 3$, i.e. $q^l=1$
and $q^n\neq 1$ for $n<l$. In this case, the center of $U_q\left(
sl_2\right) $ is generated by $C_q$ and $Z_0$, where $Z_0$ is the
commutative subalgebra of $U_q\left( sl_2\right) $ generated by the elements 
$x,y,z,z^{-1}$ with 
\begin{eqnarray*}
x &=&\left( \left( q-q^{-1}\right) e\right) ^{\overline{l}} \\
y &=&\left( \left( q-q^{-1}\right) f\right) ^{\overline{l}} \\
z &=&k^{\overline{l}} \\
z^{-1} &=&k^{-\overline{l}}
\end{eqnarray*}
and 
\[
\overline{l}=l 
\]
for $l$ odd, and 
\[
\overline{l}=\frac l2 
\]
for $l$ even. For a more detailed introduction to $U_q\left( sl_2\right) $
we refer the reader to \cite{CP} or \cite{KS}.

The quantum coadjoint action (see \cite{DK}, \cite{DKP}) on $U_q\left(
sl_2\right) $ is introduced as follows: Define derivations $\underline{e},\ 
\underline{f},\ \underline{k},\ \underline{k}^{-1}$ on $U_q\left(
sl_2\right) $ by 
\begin{eqnarray*}
\underline{e}\left( a\right) &=&\left[ \frac{e^{\overline{l}}}{\left[ 
\overline{l}\right] _q!},a\right] \\
\underline{f}\left( a\right) &=&\left[ \frac{f^{\overline{l}}}{\left[ 
\overline{l}\right] _q!},a\right] \\
\underline{k}\left( a\right) &=&\left[ \frac{k^{\overline{l}}}{\left[ 
\overline{l}\right] _q!},a\right] \\
\underline{k}^{-1}\left( a\right) &=&\left[ \frac{\left( k^{-1}\right) ^{%
\overline{l}}}{\left[ \overline{l}\right] _q!},a\right]
\end{eqnarray*}
for $a\in U_q\left( sl_2\right) $. Remember that the $q$-factorial $\left[
n\right] _q!$ is given by 
\[
\left[ n\right] _q!=\left[ n\right] _q\left[ n-1\right] _q...\left[ 1\right]
_q 
\]
These derivations stay well defined in the limit $q^l\rightarrow 1$ and can,
alternatively, be defined by 
\begin{eqnarray*}
\underline{e}\left( e\right) &=&0 \\
\underline{e}\left( f\right) &=&\frac{kq-k^{-1}q^{-1}}{q-q^{-1}}\ \frac{%
e^{l-1}}{\left[ l-1\right] !} \\
\underline{e}\left( k\right) &=&-l^{-1}xk \\
\underline{e}\left( k^{-1}\right) &=&l^{-1}xk^{-1}
\end{eqnarray*}
and 
\begin{eqnarray*}
\underline{f}\left( e\right) &=&-\frac{f^{l-1}}{\left[ l-1\right] !}\ \frac{%
kq-k^{-1}q^{-1}}{q-q^{-1}} \\
\underline{f}\left( f\right) &=&0 \\
\underline{f}\left( k\right) &=&l^{-1}yk \\
\underline{f}\left( k^{-1}\right) &=&-l^{-1}yk^{-1}
\end{eqnarray*}
We only give this alternative definition for $\underline{e},\ \underline{f}$%
, here, since these will serve as the generators of the quantum coadjoint
action (see below). The action of $\underline{e},\ \underline{f},\ 
\underline{k},\ \underline{k}^{-1}$ on the subalgebra $Z_0$ is given by 
\begin{eqnarray*}
\underline{e}\left( x\right) &=&0 \\
\underline{e}\left( y\right) &=&z-z^{-1} \\
\underline{e}\left( z\right) &=&-xz \\
\underline{e}\left( z^{-1}\right) &=&-xz^{-1}
\end{eqnarray*}
respectively 
\begin{eqnarray*}
\underline{f}\left( x\right) &=&-\left( z-z^{-1}\right) \\
\underline{f}\left( y\right) &=&0 \\
\underline{f}\left( z\right) &=&yz \\
\underline{f}\left( z^{-1}\right) &=&-yz^{-1}
\end{eqnarray*}
respectively 
\begin{eqnarray*}
\underline{k}\left( x\right) &=&xz \\
\underline{k}\left( y\right) &=&-yz \\
\underline{k}\left( z\right) &=&\underline{k}\left( z^{-1}\right) =0
\end{eqnarray*}
and 
\begin{eqnarray*}
\underline{k}^{-1}\left( x\right) &=&xz^{-1} \\
\underline{k}^{-1}\left( y\right) &=&-yz^{-1} \\
\underline{k}^{-1}\left( z\right) &=&\underline{k}^{-1}\left( z^{-1}\right)
=0
\end{eqnarray*}
Exponentiating $\underline{e}$ and $\underline{f}$ yields automorphisms of $%
U_q\left( sl_2\right) $. We denote by $\mathcal{G}$ the (infinite
dimensional) group of automorphisms of $U_q\left( sl_2\right) $ generated by
the exponentials of $\underline{e}$ and $\underline{f}$ and by $\mathcal{L}%
\left( \mathcal{G}\right) $ the Lie algebra of $\mathcal{G}$.

Now, let $\widehat{\mathcal{G}}$ be the following group: As elements of $%
\widehat{\mathcal{G}}$ we take formal expressions $\varphi \left( .\right) $
which can be evaluated at any Hopf algebra $H$ in the class of Hopf algebras
which carry a quantum coadjoint action. Denote by $\mathcal{G}_H$ the
corresponding group of automorphisms, i.e. 
\[
\mathcal{G}_{U_q\left( sl_2\right) }=\mathcal{G} 
\]
We assume that after evaluation 
\[
\varphi \left( H\right) \in \mathcal{G}_H 
\]
Define the group law of $\widehat{\mathcal{G}}$ by 
\[
\left( \varphi _1\cdot \varphi _2\right) \left( H\right) =\varphi _1\left(
H\right) \cdot \varphi _2\left( H\right) 
\]
where on the right hand side the multiplication is taken in $\mathcal{G}_H$.
It is easily checked by calculation that this together with the definition
of a unit $1$ and an inverse $\left( .\right) ^{-1}$ of $\widehat{\mathcal{G}%
}$ by 
\[
1\left( H\right) =1_{\mathcal{G}_H} 
\]
and 
\[
\varphi ^{-1}\left( H\right) =\left( \varphi \left( H\right) \right) ^{-1} 
\]
gives $\widehat{\mathcal{G}}$ the structure of a group.

Obviously, $\widehat{\mathcal{G}}$ generalizes the Ihara algebra in
precisely the same way in which the quantum coadjoint action generalizes the
classical coadjoint action and the Kirillov bracket. In this sense, the
quantum coadjoint action seems to be a concrete realization of a universal
structure related to a quantum counterpart of the motivic Galois group. We
will see in the next section that one can establish this claim for $%
U_q\left( sl_2\right) $ in a precise way by studying the Lie algebra $%
\mathcal{L}\left( \mathcal{G}\right) $ in more detail.

\bigskip

Let us close this section by shortly mentioning the role played by $%
U_q\left( sl_2\right) $ in string theory: For a stack of $k$ flat NS5 branes
the background is completely determined by vanishing R-R fields and (see 
\cite{BS}, \cite{CHS}, \cite{FGP}, \cite{Rey}) 
\begin{eqnarray*}
ds^2 &=&\eta _{\mu \nu }dx^\mu dx^\nu +e^{-2\phi }\left(
dr^2+r^2ds_3^2\right) \\
e^{-2\phi } &=&e^{-2\phi _0}\left( 1+\frac k{r^2}\right) \\
H &=&dB=-kd\Omega _3
\end{eqnarray*}
where $\mu ,\nu =0,1,...,5$ are directions tangent to the NS5 brane. Here, $%
ds_3$ and $d\Omega _3$ denote the line element and volume form,
respectively, on the $S^3$. So, in the transversal geometry of the flat NS5
there is always contained an $S^3$. Strings on the transversal $S^3$ can
always be described by a super-extension of the $SU(2)$-WZW model at level $%
k $. One can further show (see the cited literature) that the fermionic
degrees of freedom can be decoupled and one remains with the usual $SU(2)$%
-WZW model and a renormalization of the level $k$ by 
\[
k\mapsto k+2 
\]
It has further been shown in \cite{ARS} that the world-volume geometry of $D$%
-branes in the $SU(2)$-WZW model at level $k$ is described by the $q$%
-deformed fuzzy sphere of \cite{GMS}, i.e. by the representation theory of $%
U_q\left( sl_2\right) $ with 
\[
q=e^{\frac{2\pi i}{k+2}} 
\]
In the limit $k\rightarrow \infty $ one retains the usual fuzzy sphere of 
\cite{Mad} which is completely determined by the representation theory of $%
su(2)$.

In \cite{Kon 1999} and \cite{KoSo} a far reaching program was initiated to
establish an action of (a quotient of)\ the motivic Galois group on the
extended moduli space of (topological) string theory. It is tempting to ask
for a similar approach in the much simpler case of the $SU(2)$-WZW model.
The suggestion made in \cite{Kon 1999} to study an action of the motivic
Galois group on the center of $U\left( su(2)\right) $ (more precisely, on
the Duflo isomorphism of $su(2)$, see above) is very much related to trying
to find such an action for the $SU(2)$-WZW model at level $k\rightarrow
\infty $. As we have mentioned already, the action one finds in this way is
trivial. Now, $k\rightarrow \infty $ is an unphysical limit (infinite stack
of NS5 branes). The approach we follow, here, to study the center of the
quantum algebra $U_q\left( sl_2\right) $ at roots of unity, instead, can
from the string theoretic side be seen as returning to the physically more
realistic case of finite level $k$.

\bigskip

\section{The construction}

As is well known, the Lie algebra $\mathcal{L}\left( \mathcal{G}\right) $ is
infinite dimensional. Concretely, this means that the commutator $\left[ 
\underline{e},\underline{f}\right] $ does not close but is the starting
point for the definition of infinitely many new basis elements of the Lie
algebra. Here, and in the sequel, we mean by a commutator like $\left[ 
\underline{e},\underline{f}\right] $ the commutator in the sense of
derivations, i.e. 
\[
\left[ \underline{e},\underline{f}\right] =\underline{e}\circ \underline{f}-%
\underline{f}\circ \underline{e} 
\]
where $\circ $ denotes the successive application of the derivations. We
introduce the following notation: 
\begin{eqnarray*}
L &=&\underline{k}+\underline{k}^{-1} \\
e_0 &=&\underline{e} \\
f_0 &=&\underline{f}
\end{eqnarray*}
One checks that 
\[
\left[ \underline{e},\underline{f}\right] =L 
\]
We further define for all $N\in \Bbb{N}$ 
\begin{eqnarray*}
e_{N+1} &=&\left[ L,e_N\right] \\
f_{N+1} &=&\left[ L,f_N\right]
\end{eqnarray*}

\bigskip

\begin{lemma}
On the subalgebra $Z_0$ we have 
\begin{eqnarray*}
e_N\left( x\right) =Nx^2\left( z+z^{-1}\right) ^{N-1}\left( z-z^{-1}\right)
\\
e_N\left( y\right) =\left( z+z^{-1}\right) ^N\left( z-z^{-1}\right)
-Nxy\left( z+z^{-1}\right) ^{N-1}\left( z-z^{-1}\right) \\
e_N\left( z+z^{-1}\right) =\left( -x\right) \left( z+z^{-1}\right) ^N\left(
z-z^{-1}\right)
\end{eqnarray*}
and 
\begin{eqnarray*}
f_N\left( x\right) =\left( -1\right) ^{N-1}\left( z+z^{-1}\right) ^N\left(
z-z^{-1}\right) +\left( -1\right) ^NNxy\left( z+z^{-1}\right) ^{N-1}\left(
z-z^{-1}\right) \\
f_N\left( y\right) =\left( -1\right) ^{N-1}Ny^2\left( z+z^{-1}\right)
^{N-1}\left( z-z^{-1}\right) \\
f_N\left( z+z^{-1}\right) =\left( -1\right) ^Ny\left( z+z^{-1}\right)
^N\left( z-z^{-1}\right)
\end{eqnarray*}
\end{lemma}

\proof%
By calculation. 
\endproof%

\bigskip

For convenience, we have chosen the coordinate $z+z^{-1}$ instead of $z$
(observe that $z$ and $z^{-1}$ are not independent, i.e. we do not have to
take the complementary coordinate $z-z^{-1}$, in addition).

\bigskip

Let $\mathcal{A}$ be the algebra of polynomials in $z$ and $z^{-1}$.
Defining 
\[
L_0=z\frac d{dz} 
\]
and 
\[
\overline{P}_{2k+1}=z^{2k+1} 
\]
one checks that for arbitrary $\psi \in \mathcal{A}$ we have 
\begin{eqnarray*}
&&\left[ L_0,\overline{P}_{2k+1}\right] \psi \\
&=&L_0\overline{P}_{2k+1}\psi -\overline{P}_{2k+1}L_0\psi \\
&=&\left( z\frac d{dz}\overline{P}_{2k+1}\right) \psi +z\overline{P}%
_{2k+1}\left( \frac d{dz}\psi \right) -z\overline{P}_{2k+1}\left( \frac
d{dz}\psi \right) \\
&=&\left( 2k+1\right) \overline{P}_{2k+1}\psi
\end{eqnarray*}
i.e. 
\[
\left[ L_0,\overline{P}_{2k+1}\right] =\left( 2k+1\right) \overline{P}%
_{2k+1} 
\]
and $L_0$ and the $\overline{P}_{2k+1}$ give a representation of the algebra
defined by (\ref{2}) and (\ref{3}). Denote this representation by $\varrho $.

Let $I$ and $A$ be the following two linear operators on $\mathcal{A}$: 
\[
I\left( \psi \right) =z\psi 
\]
and 
\[
A\left( z\right) =z+z^{-1} 
\]
for $\psi \in \mathcal{A}$ and extension of $A$ to $\mathcal{A}$ by
requiring conservation of products and inversion.

Obviously, $I$ commutes with the $\overline{P}_{2k+1}$. We can use $I$ to
enlarge $\varrho $ by considering the additional commutators 
\[
\left[ L_0,I\overline{P}_{2k+1}\right] =\left( 2k+2\right) I\overline{P}%
_{2k+1} 
\]
i.e. the resulting algebra has the form 
\[
\left[ L_0,\overline{P}_n\right] =n\overline{P}_n 
\]
where now $n\in \Bbb{N}$. Finally, we use $A$ to apply a coordinate
transformation to the operator $\overline{P}_n$: 
\[
\overline{P}_n\mapsto \overline{P}_n\circ A 
\]
We get the resulting commutators 
\begin{equation}
\left[ L_0,P_n\right] =nP_{n-1}\left( z-z^{-1}\right)  \label{6}
\end{equation}
where 
\[
P_n=\overline{P}_n\circ A 
\]
i.e. 
\[
P_n\left( z\right) =\left( z+z^{-1}\right) ^n 
\]
We call the Lie algebra defined by (\ref{6}) $\widehat{\varrho }$. As
follows from the construction, $\widehat{\varrho }$ is directly induced by a
coordinate transformation on one of the operators and a trivial enlargement
by the operator $I$ from the representation $\varrho $.

\bigskip

We can now rewrite the result of the previous lemma as: 
\begin{eqnarray*}
e_N\left( z+z^{-1}\right) &=&\left( -x\right) \left( z+z^{-1}\right) \left[
L_0,P_N\right] \left( z\right) \\
e_N\left( x\right) &=&Nx^2\left[ L_0,P_N\right] \left( z\right) \\
e_N\left( y\right) &=&\left( z+z^{-1}-Nxy\right) \left[ L_0,P_N\right]
\left( z\right)
\end{eqnarray*}
and 
\begin{eqnarray*}
f_N\left( z+z^{-1}\right) &=&\left( -1\right) ^{N-1}\left( -y\right) \left(
z+z^{-1}\right) \left[ L_0,P_N\right] \left( z\right) \\
f_N\left( x\right) &=&\left( -1\right) ^{N-1}\left( z+z^{-1}-Nxy\right)
\left[ L_0,P_N\right] \left( z\right) \\
f_N\left( y\right) &=&\left( -1\right) ^{N-1}Ny^2\left[ L_0,P_N\right]
\left( z\right)
\end{eqnarray*}
As we have remarked already, from the four variables $x,y,z,z^{-1}$, $z$ and 
$z^{-1}$ are not independent. Actually, only two free variables remain after
taking a quotient to implement the relation 
\begin{equation}
z+z^{-1}+xy=0  \label{7}
\end{equation}
This should be done because of the following lemma:

\bigskip

\begin{lemma}
For all $N\in \Bbb{N}$ we have 
\[
e_N\left( z+z^{-1}+xy\right) =f_N\left( z+z^{-1}+xy\right) =0 
\]
\end{lemma}

\proof%
We have 
\begin{eqnarray*}
&&e_N\left( z+z^{-1}+xy\right) \\
&=&\left( -x\right) \left( z+z^{-1}\right) ^N\left( z-z^{-1}\right)
+e_N\left( x\right) y+xe_N\left( y\right) \\
&=&\left( -x\right) \left( z+z^{-1}\right) ^N\left( z-z^{-1}\right)
+Nx^2y\left( z+z^{-1}\right) ^{N-1}\left( z-z^{-1}\right) \\
&&+x\left( z+z^{-1}\right) ^N\left( z-z^{-1}\right) -Nx^2y\left(
z+z^{-1}\right) ^{N-1}\left( z-z^{-1}\right) \\
&=&0
\end{eqnarray*}
A similar calculation proves the case of the $f_N$. 
\endproof%

\bigskip

On a more abstract level, the result of the previous lemma can be seen as a
consequence of the fact that the quantum Casimir $C_q$ is invariant under
the quantum coadjoint action, i.e. the polynomial $z+z^{-1}+xy$ is
annihilated by $\mathcal{L}\left( \mathcal{G}\right) $ (see the cited
literature, see also section 2 of \cite{Kor} for a brief overview of some
properties of the quantum coadjoint action).

Using (\ref{7}), we finally have 
\begin{eqnarray}
e_N\left( z+z^{-1}\right) &=&\left( -x\right) \left( z+z^{-1}\right) \left[
L_0,P_N\right] \left( z\right)  \label{8} \\
e_N\left( x\right) &=&Nx^2\left[ L_0,P_N\right] \left( z\right)  \nonumber \\
e_N\left( y\right) &=&-\left( N+1\right) xy\ \left[ L_0,P_N\right] \left(
z\right)  \nonumber
\end{eqnarray}
and 
\begin{eqnarray}
f_N\left( z+z^{-1}\right) &=&\left( -1\right) ^{N-1}\left( -y\right) \left(
z+z^{-1}\right) \left[ L_0,P_N\right] \left( z\right)  \label{9} \\
f_N\left( x\right) &=&\left( -1\right) ^{N-1}\left( -\left( N+1\right)
\right) xy\ \left[ L_0,P_N\right] \left( z\right)  \nonumber \\
f_N\left( y\right) &=&\left( -1\right) ^{N-1}Ny^2\left[ L_0,P_N\right]
\left( z\right)  \nonumber
\end{eqnarray}
In conclusion, the operators $e_N,f_N$ give a three dimensional - using
variables $x,y,z$ - realization of the algebra $\widehat{\varrho }$. The
nontrivial action of these operators on the variable $z$ is always given by
a commutator in $\widehat{\varrho }$. The additional polynomial in $%
x,y,z,z^{-1}$ in front of the commutator is completely determined - up to a
numerical factor - by the trivial rule that $e_N$ multiplies the argument by 
$x$ and $f_N$ multiplies the argument by $y$. E.g. $e_N\left(
z+z^{-1}\right) $ receives the factor $x\left( z+z^{-1}\right) $ while $%
f_N\left( x\right) $ receives the factor $xy$. One proves that the numerical
coefficients are completely determined by the requirement (\ref{7}), then.

As an immediate consequence of (\ref{8}) and (\ref{9}), the commutators $%
\left[ e_N,e_M\right] $, $\left[ e_N,f_M\right] $, etc. are determined by
the higher commutators in $\widehat{\varrho }$ together with an extension of
the multiplication rule for the coefficients. E.g. $\left[ e_N,e_M\right]
\left( y\right) $ means that the coefficient polynomial is $x^2y$ since we
have two factors $e_N,e_M$ and, hence, a multiplication by $x^2$. In
consequence, we have the following result:

\bigskip

\begin{lemma}
The complete algebra $\mathcal{L}\left( \mathcal{G}\right) $ is induced, in
the way described above, by a representation of the solvable quotient of the
motivic Galois group given by (\ref{2}) and (\ref{3}).
\end{lemma}

\bigskip

The nontriviality of the quantum coadjoint action immediately shows that the
quotient of the motivic Galois group is realized nontrivially on $U_q\left(
sl_2\right) $.

\bigskip

\begin{remark}
The coefficient factors received by the rule that $e_N$ multiplies by $x$
and $f_N$ by $y$ strongly remind one of how one would introduce an affine
version of the algebra given by (\ref{2}) and (\ref{3}). In the case of the
classical finite dimensional Lie algebras, the theorem of Kazhdan-Lusztig
relates the affine version of the Lie algebra to a quantum deformation of
the universal envelope of the Lie algebra. A corresponding theorem for the
infinite dimensional case is not proved but we will see below that, indeed,
we should see $\mathcal{L}\left( \mathcal{G}\right) $ as a realization of a
quantum counterpart of the motivic Galois group. This is also in accordance
with the construction of $\widehat{\mathcal{G}}$ above as a quantum analogue
of the Ihara algebra. The fact that in the case of $U_q\left( sl_2\right) $
we mainly - up to a simple multiplication table - see the classical motivic
Galois group should be related to the semi-rigidity of $U_q\left(
sl_2\right) $ (i.e. one can always transform away deformations of the
coproduct) which makes the Hopf algebra cohomology of $U_q\left( sl_2\right) 
$ very close to Hochschild cohomology.
\end{remark}

\bigskip

In the next section, we will give a physics motivated ``explanation'' for
the above results which also points towards their possible generalization.

\bigskip

\section{The physics behind}

In \cite{Kon 1999}, \cite{KoSo} the action of $GT$ on Hochschild cohomology
is considered. If one passes from Hochschild- to Hopf algebra cohomology,
one expects a doubling of this $GT$ action with precisely the compatibility
relations installed which lead to the definition of the self-dual,
noncommutative, and noncocommutative Hopf algebra $\mathcal{H}_{GT}$ (see 
\cite{Sch}).

As we have mentioned in section 2, $U_q\left( sl_2\right) $ appears in
string theory as describing the world-volume geometry of $D$-branes in the $%
SU(2)$-WZW model. We now make

\bigskip

\textbf{Assumption 1:}

The generators of a suitable moduli space of the boundary $SU(2)$-WZW model
should be given by the Hopf algebra cohomology of $U_q\left( sl_2\right) $.

\bigskip

With this assumption, we can conclude that the deformation theory of the
boundary $SU(2)$-WZW model should have an action of $\mathcal{H}_{GT}$ as a
symmetry.

It is one of the general believes in the theory of the quantum coadjoint
action (see e.g. \cite{CP}) that - roughly speaking - the following holds
true:

\bigskip

\textbf{Assumption 2:}

The representation theory of a Hopf algebra with quantum coadjoint action is
determined by the quantum coadjoint action.

\bigskip

In physics terminology, this means especially that in the $SU(2)$-WZW model
the operator product expansion should be determined by knowing $\mathcal{G}$.

If we now make Assumption 3, below, we can conclude that the quantum
coadjoint action should give a realization of $\mathcal{H}_{GT}$. In the
semi-rigid case of $U_q\left( sl_2\right) $ - where we do not really have
the doubling of $GT$ to $\mathcal{H}_{GT}$ - this means that the quantum
coadjoint action should give a realization of the quotient of the motivic
Galois group which is supposed to be equivalent to $GT$. This is precisely
what we have shown in the preceding section.

\bigskip

\textbf{Assumption 3: }

The $SU(2)$-WZW model should be \textit{formal} in the sense that its
deformation theory should agree with its algebra of observables.

\bigskip

This formality assumption (the cohomology of the algebraic structure should
agree with the algebraic structure itself) is the essential input to pass
from $\mathcal{H}_{GT}$ to the quantum coadjoint action. We believe that
this is a key element to generalize our approach: Quantum field theories
which have such a formality property should give a realization of $\mathcal{H%
}_{GT}$ in terms of a generalization of the quantum coadjoint action on $%
U_q\left( sl_2\right) $.

There are indications that in many cases this formality assumption holds
true in quantum field theory: For 2d CFTs the deformations described by the
WDVV-equations (see \cite{DVV}, \cite{Wit 1991b}) are, indeed, in one to one
correspondence to the observables of the theory. If one turns on background
fields or introduces NS-branes in string theory, the deformation theory
becomes much more complicated (see the fundamental work \cite{HM}). We have
suggested in \cite{GS} a universal envelope for the BRST-complex, the
cohomology of which might describe the full deformation theory with
background fields and NS-branes. On the level of this universal envelope we,
again, expect a formality property to hold as a consequence of ultrarigidity
(see \cite{Sch} for this notion).

So, Assumption 3 might well be a general property of quantum field theory
but in more complicated models it might only hold true if one includes all
the allowed background fields into the model. Turned the other way around,
Assumption 3 could serve as a guideline to search for necessary background
fields to include in order to satisfy this principle. If such a view holds
true, our construction might be an example of a general link between $%
\mathcal{H}_{GT}$ as a quantum counterpart of the motivic Galois group and
its representation theoretic realization in quantum field theory.

\bigskip

Let us conclude this section by making three remarks on how we think one
could concretely start to generalize our approach to more complicated
backgrounds in string theory:

\begin{itemize}
\item  One should try to exploit the link between $D$-brane world-volume
geometry in the $SU(2)$-WZW model and NS5-brane backgrounds to get knowledge
about quantum motivic structures in the case of more general NS5-brane
backgrounds and for the case of little string theory (see also the remarks
in \cite{GS} on this topic).

\item  In \cite{Bez} a link between cohomology of tilting modules over
quantum groups at roots of unity and derived categories of coherent sheaves
has been shown. This might serve as a starting point to study quantum
motivic symmetries on these derived categories and could therefore be highly
relevant for the homological mirror symmetry program (\cite{Kon 1994}).

\item  Last but not least, the quantum coadjoint action on $U_q\left(
sl_2\right) $ has been shown to relate to a hidden symmetry of the
six-vertex model (see \cite{DFM}, \cite{FM 2000}, \cite{FM 2001}, \cite{Kor}%
). Can one generalize this approach to the melting crystal models appearing
in topological string theory (see \cite{ORV}, \cite{INOV})? This might be a
very direct way to establish a large hidden symmetry of a quantum motivic
type in topological string theory.
\end{itemize}

\bigskip

\section{Conclusion}

We have shown that the Lie algebra $\mathcal{L}\left( \mathcal{G}\right) $
of the quantum coadjoint action on $U_q\left( sl_2\right) $ is induced by a
representation of the solvable quotient of the motivic Galois group given by
(\ref{2}) and (\ref{3}). We have also given an argument why we think that
our construction might relate to a general formality property of quantum
field theories and pointed out different directions how to achieve such a
generalization.

We plan to study the physical implications of this link between the quantum
coadjoint action and the motivic Galois group in more detail in the near
future.

\end{document}